\documentclass[journal]{IEEEtran}

\usepackage{amsmath,amssymb,amsfonts}
\usepackage{cite}
\usepackage{graphicx}
\usepackage{textcomp}
\usepackage{xcolor}
\usepackage{color,soul}
\usepackage{multirow}
\usepackage[caption=false,font=footnotesize]{subfig}  
\usepackage{url}
\usepackage{amsthm}
\usepackage{algorithm,algorithmic}
\usepackage{fancyhdr}
\usepackage{lastpage}
\usepackage{comment}

\begin{document}

\IEEEaftertitletext{\vspace{-2.5\baselineskip}}

\title{Admittance Identification of Grid-Forming Inverters Using Time and Frequency-Domain Techniques\vspace{-1mm}}

\author{
\IEEEauthorblockN{\textbf{Andres Intriago}\IEEEauthorrefmark{1}, \textbf{Alexandros Paspatis}\IEEEauthorrefmark{2}, \textbf{Francesco Liberati}\IEEEauthorrefmark{3}, \textbf{Charalambos Konstantinou}\IEEEauthorrefmark{1}}

\IEEEauthorblockA{\IEEEauthorrefmark{1}CEMSE Division, King Abdullah University of Science and Technology (KAUST)\\
\IEEEauthorrefmark{2}Department of Engineering, Manchester Metropolitan University, Manchester, United Kingdom\\
\IEEEauthorrefmark{3}Department of Computer, Control and Management Eng. Antonio Ruberti, University of Rome La Sapienza}

}

\maketitle

\begin{abstract}

The increasing integration of inverter-based resources (IBRs) into the power grid introduces new challenges, requiring detailed electromagnetic transient (EMT) studies to analyze system interactions. Despite these needs, access to the internal firmware of power electronic devices remains restricted due to stringent nondisclosure agreements enforced by manufacturers. To address this, we explore three system identification techniques: sweep frequency response analysis (SFRA), step excitation method (SEM), and eigensystem realization algorithm (ERA). SFRA employs sinusoidal signals of varying frequencies to measure the system's frequency response, while SEM and ERA utilize step functions to derive time-domain responses and transform them into Laplace-domain transfer functions. All three approaches are shown to provide consistent results in identifying the $dq$ admittance of grid-forming inverters (GFM) over a frequency range of $1$ Hz to $100$ Hz.
\end{abstract}

\vspace{-4mm}

\section{Introduction}
\vspace{-1mm}




The increasing penetration of inverter-based resources (IBRs) introduces new stability challenges in power systems, as inverters, unlike synchronous generators (SGs), lack the physical inertia needed to counteract power imbalances. This deficiency makes it essential to develop practical methods for stability analysis to maintain reliable grid operations \cite{Cheng_2022}. One key manifestation of such stability issues is sub-synchronous oscillations, which exhibit characteristics distinct from those in conventional power systems \cite{Bialek2020}. To address these challenges, a comprehensive framework for assessing the integration of IBRs is crucial. Two primary methods are commonly employed to analyze power system dynamics: state-space analysis and impedance-based analysis. Each technique offers unique advantages and is chosen based on specific requirements and the context of the stability problem.

State-space techniques assess the system's stability by performing a small-signal analysis of the grid network, computing eigenvalues, and mapping zeros and poles in the complex plane \cite{Zhu2021}. They provide detailed insights into the dynamics of the states allowing for a comprehensive understanding of unstable modes through sensitivity analysis \cite{Zhu2023_2}. State-space methods can be used to model traditional power systems and networks with high penetration of renewable energy resources (RES), such as RES-based microgrids (MGs) \cite{Intriago_2024}. However, state-space modeling relies on complete knowledge of the control algorithms and the differential equations that describe the dynamics of IBRs. This information is typically not available for power electronic devices, as each vendor has its control architecture, which is often not disclosed \cite{Fan_2021}. 

On the other hand, impedance-based analysis is considered an alternative to the state-space method for stability assessment \cite{Wang2014}. This approach allows for the identification of impedance and admittance using data-driven frameworks, eliminating the need for complete access to the system model \cite{Sun2011}. For impedance identification, small-signal current injections are utilized to observe the resulting voltage transients \cite{Wen2015}. Conversely, admittance identification involves injecting voltage perturbations to measure the corresponding current transients \cite{Shah2019}. These techniques extract models that capture the system's dynamics, ensuring that each state is controllable and observable from the measurement point. Thus, impedance and admittance-based methods provide stability insights comparable to those achieved through state-space analysis \cite{Zhu2023}.

This paper examines various approaches for accurately extracting the $dq$ admittance using three system identification frameworks. While all methods are effective, using the step response function alone can achieve this more efficiently compared to frequency scanning. We conduct a critical comparison of the time-domain and frequency-domain methods, evaluating factors such as input signal characteristics, limitations, benefits, and the scenarios where each technique is most applicable. 
The remainder of the paper is structured as follows. Section \ref{s:methodology} presents the methodology of the identification algorithms. Section \ref{s:comparison} discusses the scenarios in which time or frequency domain techniques are better suited. Section \ref{s:results} presents simulation results, while Section \ref{s:conclusion} concludes the article.


\vspace{-3mm}
\vspace{-2mm}
\section{Admittance Identification Techniques of IBRs}\label{s:methodology}

\vspace{-1mm}
\subsection{System Modeling}
\vspace{-1mm}

\begin{figure}[t]
    \centering
    \includegraphics[trim=0.1cm 0cm 0.05cm 0cm, clip, width=3.3in]{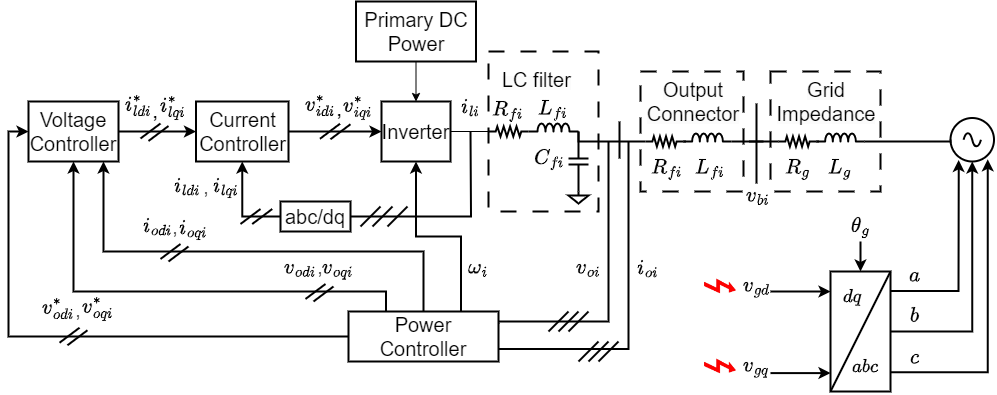}
    \vspace{-1mm}
    \caption{{{dq}-frame admittance measurement testbed.}}
    \vspace{-5mm}
    \label{fig:VCVSI_testbed_ident}
\end{figure}

Fig. \ref{fig:VCVSI_testbed_ident} presents the testbed employed to extract the $dq$ frame admittance matrix of the IBR. The controllable voltage source is connected to the IBR at the point of common coupling (PCC). Two perturbation voltages are applied one at a time to the $dq$ voltage source. The functions used to perform these perturbations depend on the identification method. The voltages are defined as $v_{gd}$, and $v_{gq}$, and converted to the $abc$-frame using $\theta_{g}$ as the angle of the grid to form a {three}-phase voltage source. The transient currents in time domain are recorded at the PCC and converted to the $dq$-frame, defined as $i_{od}$, and $i_{oq}$. The main goal of the identification process is to compute the $s$-domain outputs, specifically \( i_{od}^{(1)}(s) \), \( i_{oq}^{(1)}(s) \), \( i_{od}^{(2)}(s) \), and \( i_{oq}^{(2)}(s) \) with the superscripts indicating the number of the input. For instance, \( i_{od}^{(1)}(s) \) and \( i_{oq}^{(1)}(s) \) represent the outputs due to changes in the first input voltage defined as \( v_{gd} \), while \( i_{od}^{(2)}(s) \) and \( i_{oq}^{(2)}(s) \) represent the outputs affected by changes in the second input voltage \( v_{gq} \). The admittance matrix of the IBR  is defined with a negative sign, indicating that the current flows from the inverter to the grid.

\vspace{-3mm}
\subsection{Eigensystem Realization Algorithm (ERA)}
\begin{figure}[t]
    \centering
    \includegraphics[width=0.95\columnwidth]{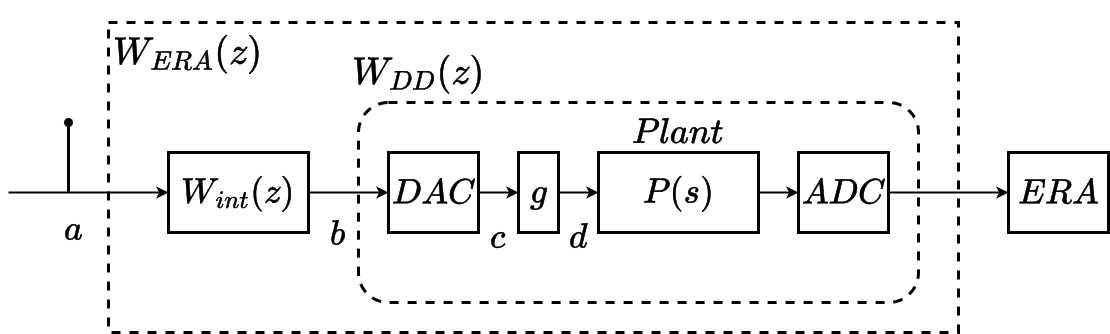}
    \vspace{-1mm}
    \caption{Reference scenario for the ERA experiment.  
    a) Discrete impulse.  
    b) Discrete step.  
    c) Continuous step.  
    d) Continuous step of amplitude $g$.}
    \label{era_scheme}
    \vspace{-6mm}
\end{figure}

\vspace{-1mm}
ERA generates reduced-rank linear input-output models by processing impulse response time series, without the need for prior system knowledge \cite{Juang1985}. However, the practical application of ERA presents a challenge, as it relies on discrete impulse signals to capture transient responses, while real-world power systems operate in continuous time. Fig. \ref{era_scheme} presents the experimental setup designed to effectively apply ERA and address this technical limitation.

To elaborate on this setup, consider an experiment where the input to the system to be identified, with transfer function $P(s)$, is a continuous-time voltage step with an amplitude of $g$ (point d). As illustrated in the diagram, the continuous-time step can be realized as the output of a zero-order digital-to-analog converter (point c), whose input is given by the discrete step (point b). The discrete step in turn can be realized as the output of a digital integrator when the input is given by the discrete impulse (point a). Therefore, the sampled step response of $P(s)$ is the same as the discrete-impulse response of the discrete-time extended system in the outer dashed box in Fig. \ref{era_scheme}. The extended system is the one which is identified by the ERA when the algorithm is fed with the step response of the plant. We denote the discrete transfer function of the extended system as $W_{ERA}(z)$ (the one in the outer dashed box in Fig. \ref{era_scheme}), which can be computed as a series connection of its subsystems, $W_{int}(z)$ and $W_{DD}(z)$. The discrete-time integrator transfer function is defined as: {$W_{int} = {z}/({z-1})$}.

\normalsize
Within the discrete-time system outlined by the inner dotted box in Fig. \ref{era_scheme}, the transfer function $W_{DD}(z)$ can be determined by ensuring that the discrete-time step response aligns with the sampled step response of the continuous-time system $gP(s)$.
\begin{equation}
\mathcal{Z}^{-1}\bigg\{W_{DD}(z)\frac{z}{z-1}\bigg\} = \mathcal{L}^{-1}\bigg\{g\frac{P(s)}{s}\bigg\},
\end{equation}
where $z/(z-1)$ is the $z$-transform of the discrete time step.
Hence, we have:
\begin{equation}
W_{DD}(z) = \frac{z-1}{z}\mathcal{Z}\Bigg\{\mathcal{L}^{-1}\bigg\{g\frac{P(s)}{s}\bigg\}\Bigg\}.
\end{equation}
As a result, $W_{ERA}(z)$ is calculated as following:
\begin{equation}\label{ERA_P/s}
W_{ERA}(z) = W_{int}(z)W_{DD}(z) = \mathcal{Z}\Bigg\{\mathcal{L}^{-1}\bigg\{g\frac{P(s)}{s}\bigg\}\Bigg\}.
\end{equation}
%

Let $W_{ERA}(s) = d2c(W_{ERA}(z))$ denote the continuous-time equivalent. From the previous equation, we have:%
\begin{equation}\label{ERA_P_s1}
d2c(W_{ERA}(z)) = g\frac{P(s)}{s},
\end{equation}
%
%
\begin{equation}\label{ERA_P_s2}
P(s) = d2c(W_{ERA}(z))\frac{s}{g}.
\end{equation}

Once \( P(s) \) has been identified from each measurement using \eqref{ERA_P_s2}, it is then stacked into each one of the elements of the admittance matrix, as shown in \cite{Fan_2021}: 
\begin{equation}
Y_{\text{ERA}} (s)= -\frac{s}{g} \begin{bmatrix}
i_{od}^{(1)}(s) & i_{od}^{(2)}(s) \\
i_{oq}^{(1)}(s) & i_{oq}^{(2)}(s)
\end{bmatrix}
\label{ERA_Admittance}
\end{equation}

\vspace{-6mm}
\subsection{Step Excitation Method (SEM)}
\vspace{-1mm}

SEM provides an estimation of the multiple-input and multiple-output transfer function, representing the admittance of the IBR. This approach transforms time-domain $dq$-frame measurements to continuous-time transfer functions. The admittance is computed by applying a step function at the $dq$ voltage source. 
SEM requires two step injections, where each perturbation is performed separately, ensuring that the step responses are generated from a single event (independent step changes). The step function is applied to each component of the $dq$ voltage source. By performing inverse Park transformation, it is possible to transform the $dq$ voltage source to the $abc$ {three}-phase voltage source, as depicted in Fig. \ref{fig:VCVSI_testbed_ident}. The {three}-phase current flowing from the inverter to the PCC is recorded and transformed to the $dq$ frame using the Park transformation. Subsequently, the effect of the perturbation at the $d$-axis of the voltage source in time domain can be defined as \( i_{od}^{(1)}(t) \) and \( i_{oq}^{(1)}(t) \), where these two datasets are employed to estimate $Y_{dd}(s)$ and $Y_{qd}(s)$. Furthermore, the voltage perturbation at the $q$-axis yields \( i_{od}^{(2)}(t) \) and \( i_{oq}^{(2)}(t) \) where these two datasets are employed to estimate $Y_{dq}(s)$ and $Y_{qq}(s)$. After estimating the transfer functions employing the \textsc{MATLAB} function \texttt{tfest}, the admittance matrix is described as follows:
\begin{equation}
\label{SEM_equations}
\begin{aligned}
Y_{dd}(s) &= \frac{i_{od}^{(1)}(s)}{v_{gd}^{(1)}(s)} \quad 
Y_{dq}(s) = \frac{i_{od}^{(2)}(s)}{v_{gq}^{(2)}(s)} \\
Y_{qd}(s) &= \frac{i_{oq}^{(1)}(s)}{v_{gd}^{(1)}(s)} \quad 
Y_{qq}(s) = \frac{i_{oq}^{(2)}(s)}{v_{gq}^{(2)}(s)}
\end{aligned}
\end{equation}
\begin{equation}
Y_{\text{SEM}} (s)= -\begin{bmatrix}
Y_{dd}(s) & Y_{dq}(s) \\
Y_{qd}(s) & Y_{qq}(s)
\end{bmatrix}
\label{step_Admittance}
\end{equation}

\vspace{-3mm}
\subsection{Sweep Frequency Response Analysis (SFRA)}
\vspace{-1mm}

SFRA is a frequency scanning technique used to study system behavior across multiple frequencies. This method extracts the $dq$ admittance of the IBR by injecting sinusoidal signals into the $dq$ voltage of the grid while the system operates in a steady state. The components of the $dq$ voltage source are excited separately, at different discrete frequencies. The currents in time domain are recorded at the PCC. They are transformed to the $dq$-frame variables defined as $i_{odq}$. Fast Fourier Transform is applied to compute the phasor form of the inputs $v_{gdq}(t)$ and the outputs $i_{odq}(t)$ at the frequency of the injected perturbation. After this step, the \textsc{MATLAB} function \texttt{tfest} is employed to estimate the transfer functions described in \eqref{step_Admittance}. \vspace{-2mm}
\begin{equation}
\label{freq_scan}
\begin{aligned}
Y_{dd}(f_{i}) &= -\frac{i_{od}^{(1)}(f_{i})}{v_{gd}^{(1)}(f_{i})} \quad 
Y_{dq}(f_{i}) = -\frac{i_{od}^{(2)}(f_{i})}{v_{gq}^{(2)}(f_{i})} \\
Y_{qd}(f_{i}) &= -\frac{i_{oq}^{(1)}(f_{i})}{v_{gd}^{(1)}(f_{i})} \quad 
Y_{qq}(f_{i}) = -\frac{i_{oq}^{(2)}(f_{i})}{v_{gq}^{(2)}(f_{i})}
\end{aligned}
\end{equation}

\vspace{-2mm}

\vspace{-4mm}
\section{Domain Selection For System Identification}\label{s:comparison}

\vspace{-1mm}
\subsection{Time-Domain Analysis}
\vspace{-1mm}


Both ERA and SEM operate using time-domain data. ERA typically uses a discrete impulse function as a perturbation signal to extract system information \cite{Juang1985}, but a step function can also be employed, considering its magnitude and Laplace transform as shown in \eqref{ERA_Admittance}. ERA is flexible in signal choice as long as the $z$-transform is known. In cases where isolated impulse injection are difficult to implement or sensor noise is present, a pseudo-random signal combined with the Observer Kalman Filter Identification can produce a denoised impulse response \cite{Brunton2019}. SEM, on the other hand, applies a step change at the $dq$ voltage source to collect time-domain data, which is processed to estimate continuous-time transfer functions using MATLAB's \texttt{tfest} function \cite{Ramakrishna2024}. Alternatively, Gaussian-distributed signals can be used to estimate the admittance matrix by adjusting the mean and standard deviation of the injected signal \cite{Fan2023}.


The advantages of using time-domain techniques like ERA and SEM  is the flexibility in signal design, i.e., the step function in both the ERA and SEM is not constrained by specific magnitude or duration. Similarly, Gaussian pulse methods allow for straightforward parameter adjustments. These techniques facilitate quick system identification, requiring only two perturbations to derive the system's admittance \cite{Ramakrishna2024}. However, ERA and SEM can be sensitive to non-ideal measurement data. When noise is present, estimating the correct system order becomes challenging, leading to risks of model overfitting or underfitting and potentially compromising system fidelity \cite{Juang1985}. Additionally, these methods may struggle to sufficiently excite the system across a wide frequency spectrum, resulting in less accurate high-frequency dynamics \cite{Fan2023}.




Despite these challenges, ERA and SEM are effective in small-scale networks due to their simplicity, computational efficiency, and stability \cite{Juang1985}. For example, ERA has been applied to identify the admittance matrix of a 200 MW Type-4 wind plant connected to the grid, demonstrating its use in assessing the stability of IBRs based on dynamic event data \cite{Wang_2022}. Meanwhile, SEM has been employed to extract the $dq$-admittance of solar PV farms, utilizing shunt current and series voltage injections for accurate identification \cite{Ramakrishna2024}.

\vspace{-4mm}
\subsection{Frequency-Domain Analysis}
\vspace{-1mm}

Chirp signal injection and SFRA are frequency scanning techniques commonly used to excite systems across a range of frequencies with sinusoidal waves. The chirp technique injects a sinusoidal signal with a frequency that gradually sweeps from low to high, maintaining a fixed, small magnitude, allowing the system's continuous frequency response to be captured in a single experiment \cite{Fan2023}. In contrast, SFRA injects sinusoidal signals at discrete frequencies, with the number of experiments determined by the selected  points \cite{Wang_2022}. Both techniques offer flexibility in designing the sinusoidal signal and allow the selection of a frequency vector tailored to analyze higher frequency dynamics, typically above $100$ Hz. Approaches such as SFRA, chirp, and sinusoidal harmonic injection provide improved frequency response accuracy due to their high signal-to-noise (SNR) ratios \cite{Wang_2022}.

The flexibility extends to the number of cycles per frequency point, ensuring data collection only after the system reaches a steady state, as premature measurements can compromise accuracy \cite{Zhu2023}. Studies show no strict constraints on the sinusoidal magnitude, although it is recommended that sinusoidal voltages remain within 5\% of the nominal value to extract admittance models, and sinusoidal currents within 10\% to measure impedance \cite{Riccobono2018}. In some cases, multiple magnitudes (e.g., 0.5\%, 5\%, and 10\%) are injected to refine accuracy, but guidelines on appropriate magnitudes remain limited \cite{Shah2019}.

Despite their advantages, frequency-domain techniques like SFRA and chirp are computationally intensive due to the need for multiple sinusoidal injections across frequency points. Each frequency point often requires more than one cycle, increasing both time and computational demands \cite{Fan2023}. These techniques are especially effective for large-scale systems due to their ability to handle sensor noise, yielding realistic results. SFRA has been used for small-signal stability analysis of the NETS-NYPS 14- and 68-bus system \cite{Singh2013, Zhu2023_2}.

\vspace{-3mm}
\section{Comparative Analysis}\label{s:results}

\vspace{-1mm}

This section presents simulation results demonstrating the effectiveness of the three techniques described in Section  \ref{s:methodology} and discussed in Section \ref{s:comparison} for extracting the $dq$ admittance matrix of the GFM. The adopted testbed setup is illustrated in Fig. \ref{fig:VCVSI_testbed_ident}, with relevant GFM modeling parameters provided in Table \ref{table_GFM_VS}. For all three frameworks, data is collected at a sampling frequency of $2.5$ kHz. In the case of ERA and SEM, the sampling duration ranges from $0.5$ to $1$ second, corresponding to the duration of the step function. For SFRA, the sampling time is determined by the frequency of the injected sinusoidal signal. Since both ERA and SEM use step functions to derive the admittance matrix, their dynamic responses in the time domain are identical. It is also assumed that the system is in a steady state at the time of the perturbation.



\begin{figure*}[t]
\vspace{-2mm}
\centering
    \subfloat[]{
        \includegraphics[width=0.2\textwidth]{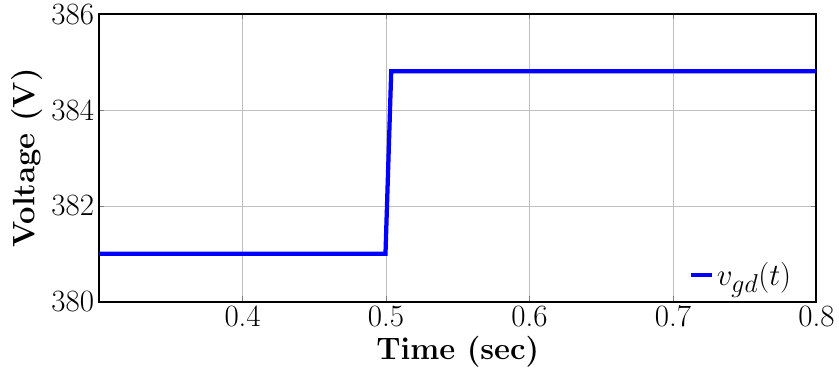}
        \label{fig:vd_on}
    } 
    \subfloat[]{
        \includegraphics[width=0.2\textwidth]{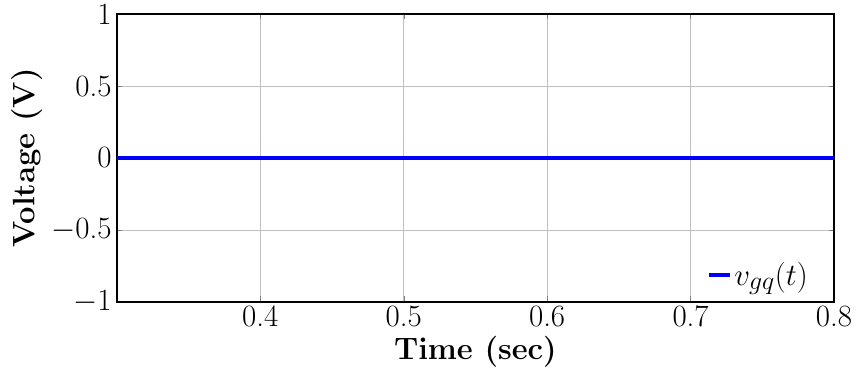}
        \label{fig:vq_off}
    } 
    \subfloat[]{
        \includegraphics[width=0.2\linewidth]{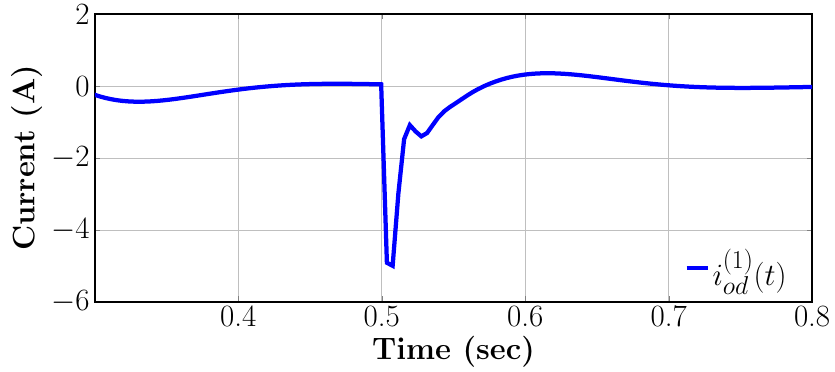}
        \label{fig:iod1}
    } 
   \subfloat[]{
        \includegraphics[width=0.2\linewidth]{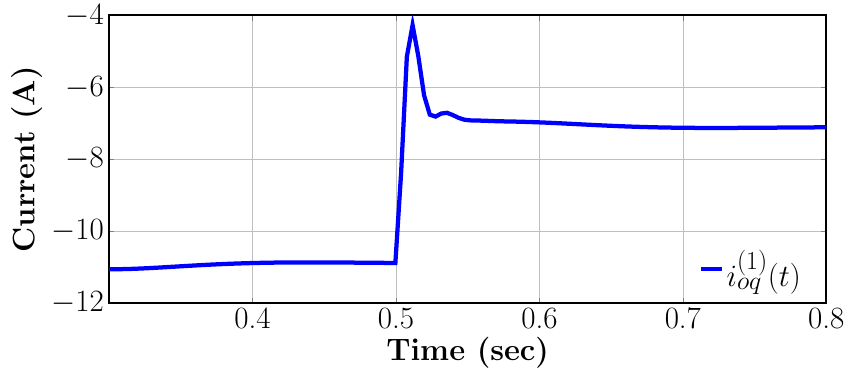}
        \label{fig:ioq1}
    } \\
\vspace{-2mm}
\caption[CR]{Step change of 1\% at $v_{gd}$ and resulting effect on $i_{od}$ and $i_{oq}$.} 
\vspace{-8mm}
\label{fig:vgd_step_change}
\end{figure*}
\begin{figure*}[t]
\centering
    \subfloat[]{
        \includegraphics[width=0.2\textwidth]{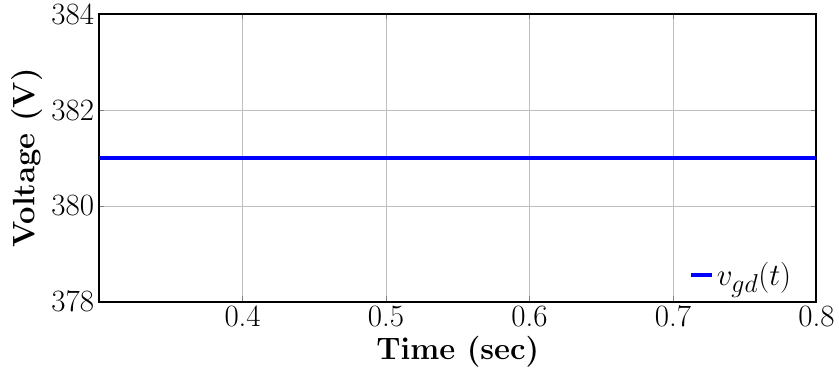}
        \label{fig:vd_off}
    } 
    \subfloat[]{
        \includegraphics[width=0.2\textwidth]{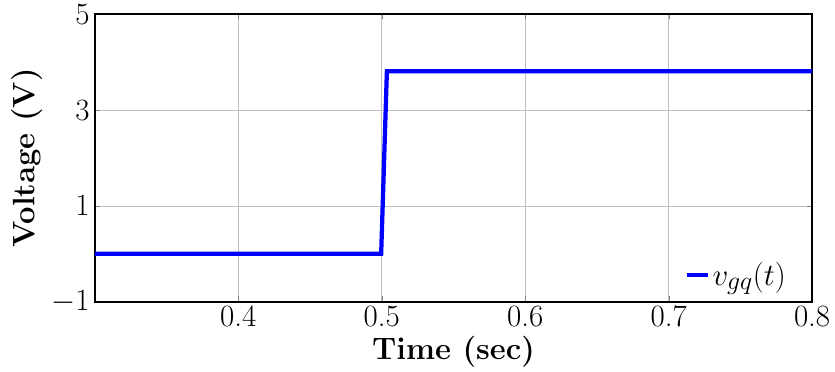}
        \label{fig:vq_on}
    } 
    \subfloat[]{
        \includegraphics[width=0.2\linewidth]{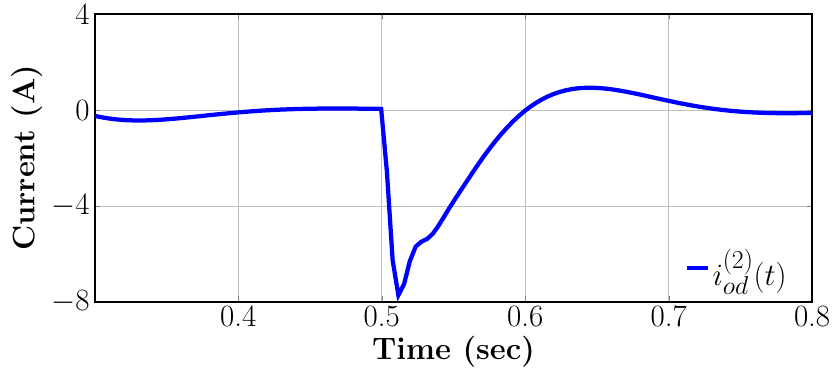}
        \label{fig:iod2}
    } 
   \subfloat[]{
        \includegraphics[width=0.2\linewidth]{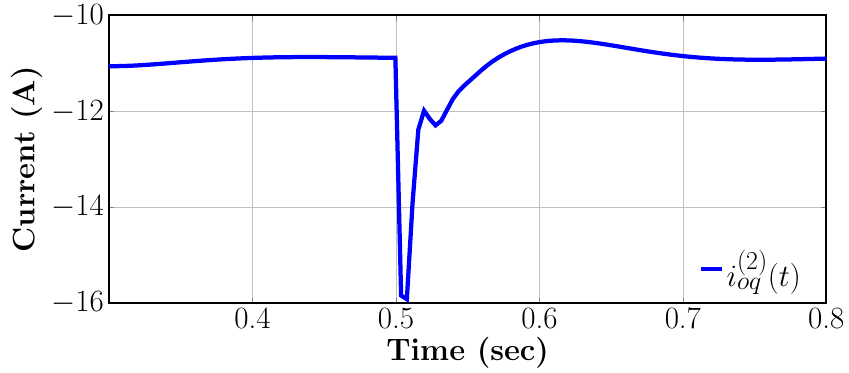}
        \label{fig:ioq2}
    } \\

\vspace{-2mm}
\caption[CR]{Step change of 1\% at $v_{gq}$ and resulting effect on $i_{od}$ and $i_{oq}$. } 
\vspace{-5mm}
\label{fig:vgq_step_change}
\end{figure*}

In ERA and SEM, a $1$\% step change is applied to $v_{gd}$, with the resulting line currents recorded by sensors and labeled as $i_{od}^{(1)}(t)$ and $i_{oq}^{(1)}(t)$. Next, a $1$\% step change is applied to $v_{gq}$, and the corresponding line currents are recorded as $i_{od}^{(2)}(t)$ and $i_{oq}^{(2)}(t)$. These responses are illustrated in Figs.~\ref{fig:vgd_step_change} and \ref{fig:vgq_step_change}.

\begin{table}[t]
\captionsetup{font=footnotesize}
\centering
\caption{Parameters of GFM and voltage source.}
\vspace{-2mm}
\begin{tabular}{||c|c|c|c||} 
\hline \hline
Parameter & Value & Parameter & Value \\
\hline \hline
$V_{ni}$ & $381$ V & $K_{PC}$ & $15$ \\
\hline
$\omega_{ni}$ & $377$ rad/s & $K_{IC}$ & $20000$ \\
\hline
$V_{DC}$ & $1000$ V & $\omega_{b}$ & $377$ rad/s \\
\hline
$m_{P}$ & $9.4 \cdot 10^{-5}$ & $F$ & $0.75$ \\
\hline
$n_{Q}$ & $1.3 \cdot 10^{-3}$ & $\omega_{c}$ & $37.7$ rad/s \\
\hline
$R_{c}$ & $0.03$ $\Omega$ & $\omega_{g}$ & $377$ rad/s \\
\hline
$L_{c}$ & $1$ mH & $R_{grid}$ & $0.23$ $\Omega$ \\
\hline
$R_{f}$ & $0.001$ $\Omega$ & $L_{grid}$ & $318$ $\mu$H \\
\hline
$L_{f}$ & $0.3$ mH & $V_{gd}$ & $380$ V \\
\hline
$C_{f}$ & $10$ $\mu$F & $V_{gq}$ & $0$ V \\
\hline
$K_{PV}$ & $0.1$ & $P_{load}$ & $12 \times 10^{3}$ W \\
\hline
$K_{IV}$ & $420$ & $Q_{load}$ & $12 \times 10^{3}$ VAR \\
\hline \hline
\end{tabular}
\label{table_GFM_VS}
\vspace{-7mm}
\end{table}

\begin{figure*}[t]
\vspace{-4mm}
\centering
    \subfloat[]{
        \includegraphics[width=0.2265\textwidth]{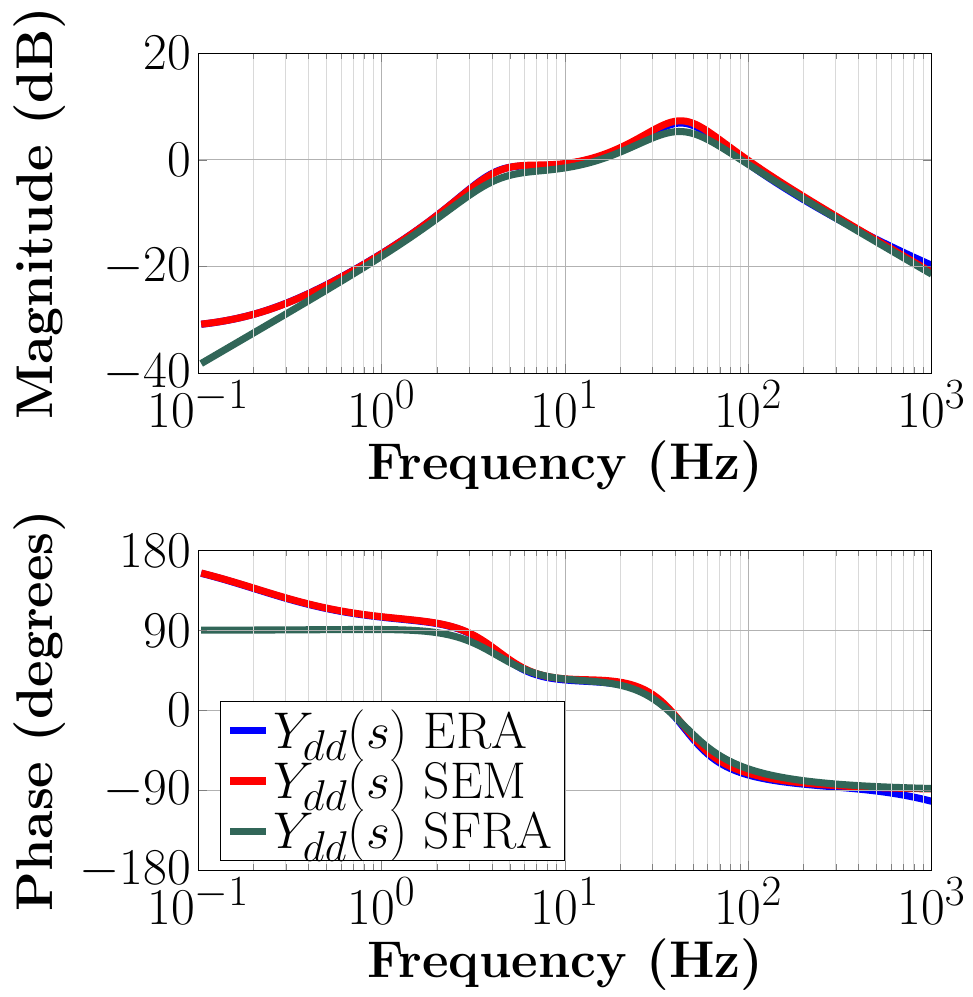}
        \label{fig:Ydd_admit}
    } 
    \subfloat[]{
        \includegraphics[width=0.2265\textwidth]{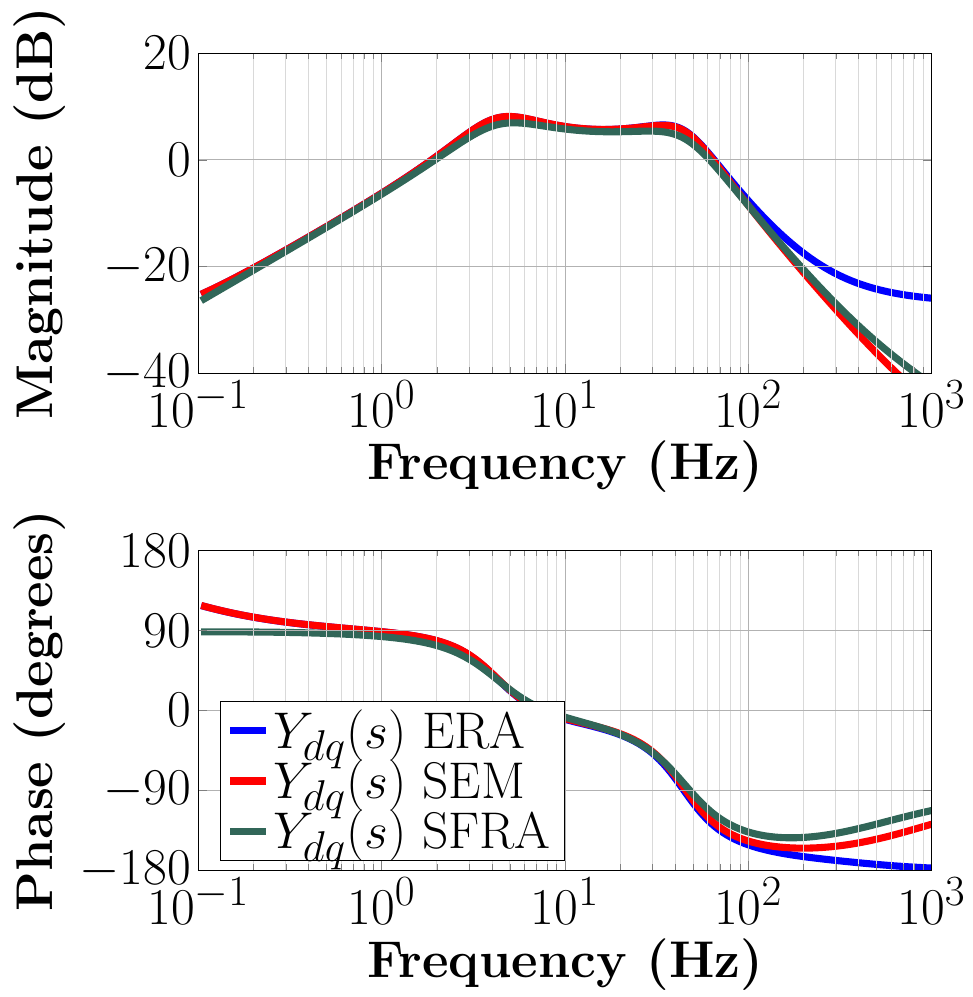}
        \label{fig:Ydq_admit}
    } 
    \subfloat[]{
        \includegraphics[width=0.2265\linewidth]{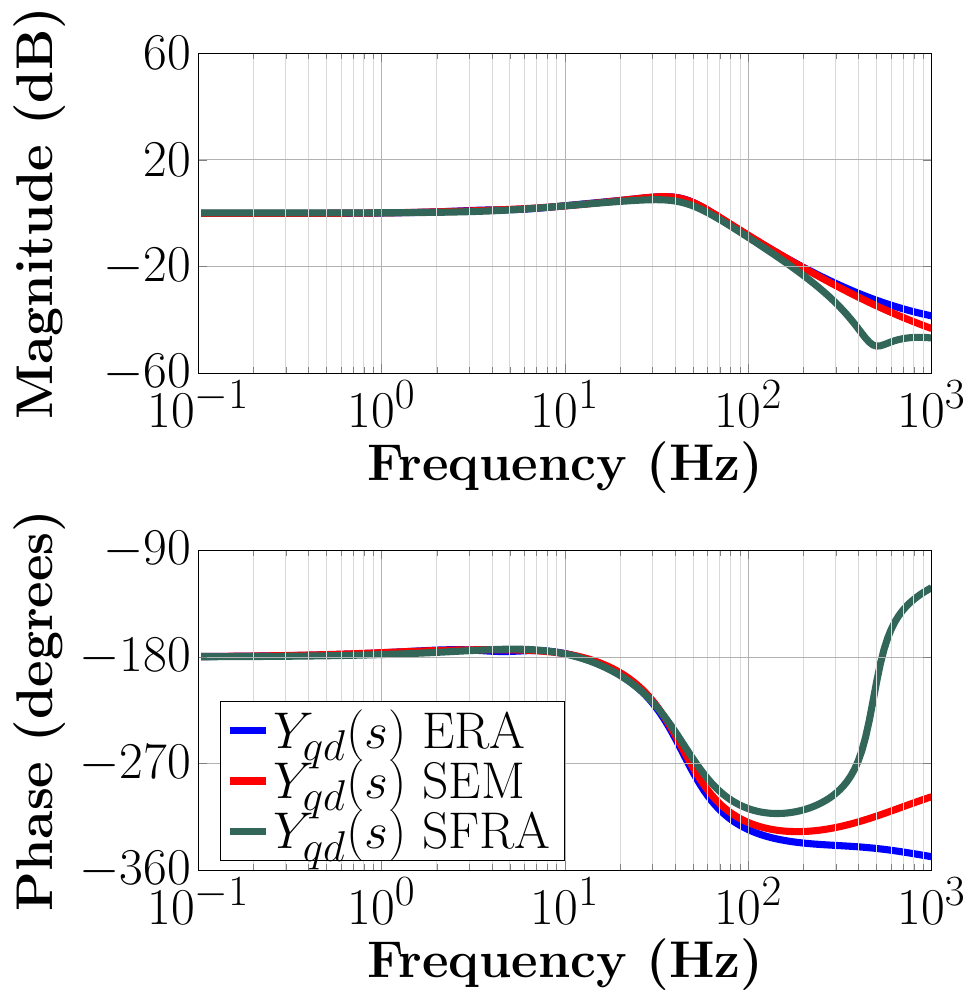}
        \label{fig:Yqd_admit}
    } 
   \subfloat[]{
        \includegraphics[width=0.2265\linewidth]{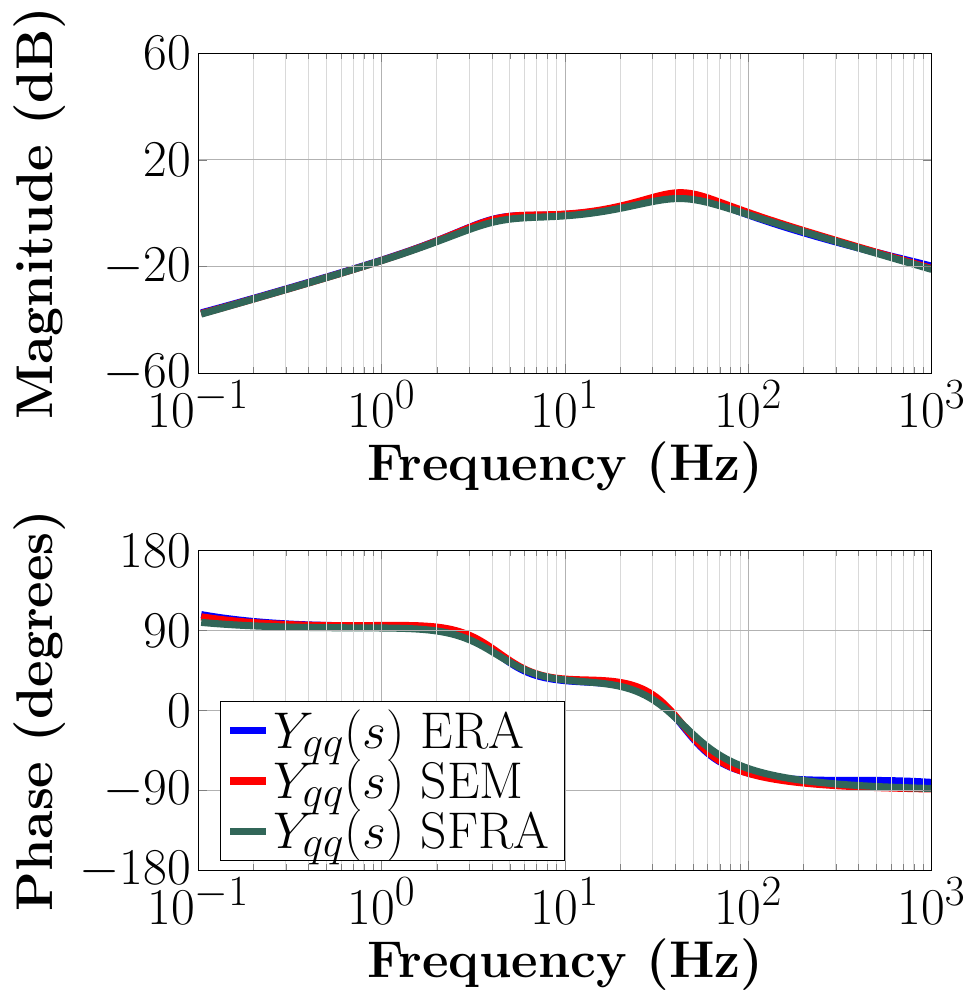}
        \label{fig:Yqq_admit}
    } \\

\vspace{-2mm}
\caption[CR]{DQ admittance extraction comparison employing the three frameworks, ERA, SEM, and SFRA.} 
\vspace{-6mm}
\label{fig:admittance_plots}
\end{figure*}

A Hankel matrix is created for ERA by stacking shifted time series of step response outputs. Truncation is required based on the system order when performing the singular value decomposition (SVD) of the Hankel matrix, as the system order is a key parameter for computing ERA \cite{Juang1985}. In this paper, we assume a black-box modeling approach in which truncation is performed iteratively, starting from the lowest order and increasing until the Bode plot of the admittance matrix approximates the results from the other two frameworks. In our case, the estimated system order is determined to be sixth order, as it best matches the other frameworks. ERA then computes the discrete matrices of the state-space model, which are used to extract the admittance transfer function in discrete time. Once this function for each measurement ($W_{ERA}$) is obtained, a transition from discrete to continuous time is made using \eqref{ERA_P_s1} and \eqref{ERA_P_s2}. This transfer function, defined as $P(s)$, represents an element of the admittance matrix in \eqref{ERA_Admittance} and is depicted in the Bode plot in Fig.~\ref{fig:admittance_plots}.

\begin{figure*}[t]
\vspace{-3mm}
\centering
    \subfloat[]{
        \includegraphics[width=0.2265\textwidth]{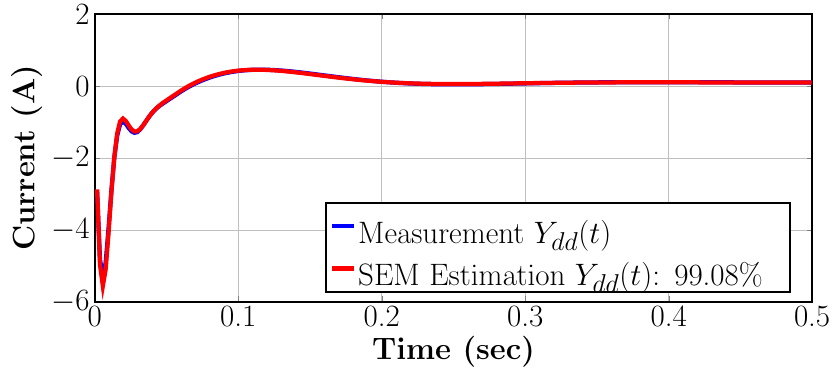}
        \label{fig:Ydd_SEM}
    } 
    \subfloat[]{
        \includegraphics[width=0.2265\textwidth]{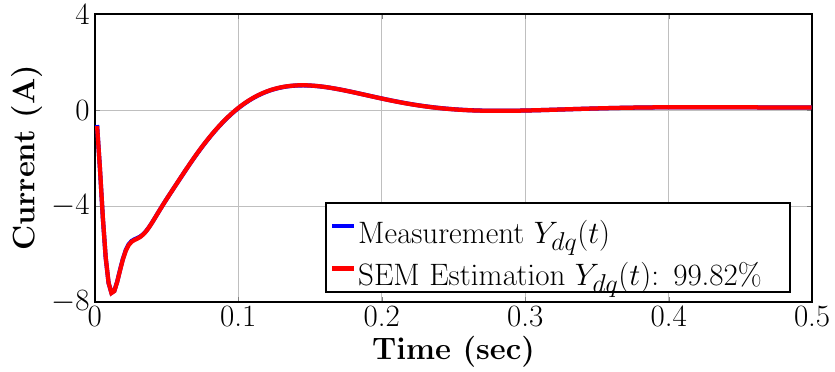}
        \label{fig:Ydq_SEM}
    } 
    \subfloat[]{
        \includegraphics[width=0.2265\linewidth]{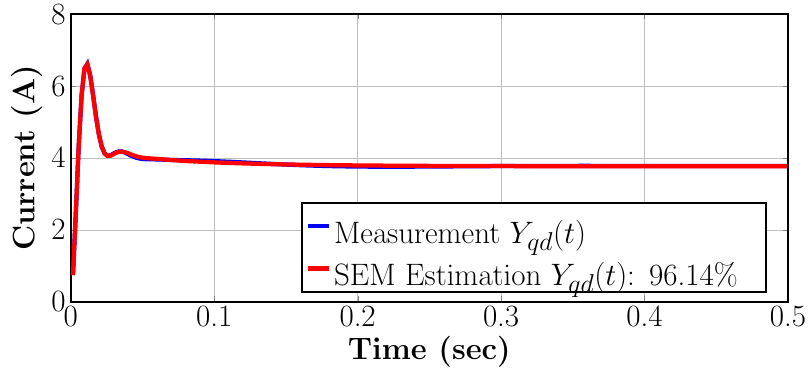}
        \label{fig:Yqd_SEM}
    } 
   \subfloat[]{
        \includegraphics[width=0.2265\linewidth]{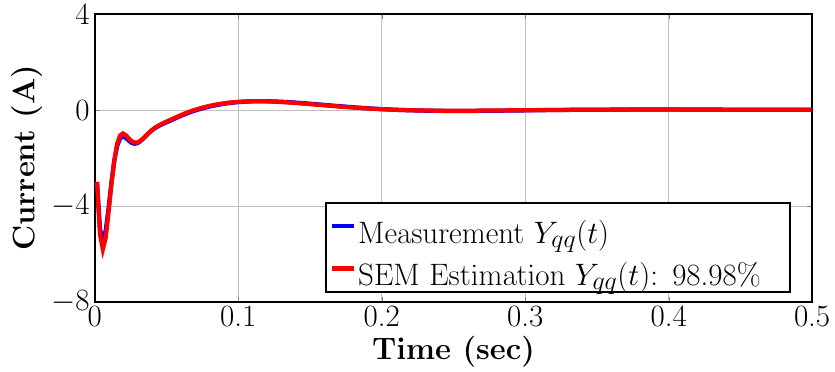}
        \label{fig:Yqq_SEM}
    } \\

\vspace{-2mm} 
\caption[CR]{Percentage of accuracy between the measurement time domain data and the estimation employing SEM.} 
\vspace{-7mm}
\label{fig:SEM_estimation}
\end{figure*}

To extract the continuous-time input-output signal in SEM, a $1$\% step change is applied separately to the $dq$ voltages of the grid. The resulting step responses are shown in Figs. \ref{fig:vgd_step_change} and \ref{fig:vgq_step_change}. The accuracy of the transfer function estimation depends heavily on the quality of the input data provided to the \texttt{tfest} algorithm. To ensure zero-mean time-series data, the DC offset, corresponding to the steady-state of the signal, has been removed. \texttt{tfest} requires the number of system poles as an input parameter. This value is determined through an iterative process, starting with a low-order model. The time-domain data is then loaded into the estimation algorithm to compute the transfer function. For this system, the number of poles is chosen to be four, as this value produces an accurate match between the system's real output and the estimated transfer function. To validate the accuracy of the estimated transfer function, a comparison is made between the model's time-domain response and the measured sensor data, as shown in Fig. \ref{fig:SEM_estimation}. The fidelity of this comparison is quantified using the normalized root mean square error (NRMSE), which is calculated by \texttt{tfest}. The NRMSE value is displayed in the legend of the simulated response comparison in Fig. \ref{fig:SEM_estimation}. 


\begin{figure*}[t]
\vspace{-2mm}
\centering
    \subfloat[]{
        \includegraphics[width=0.2265\textwidth]{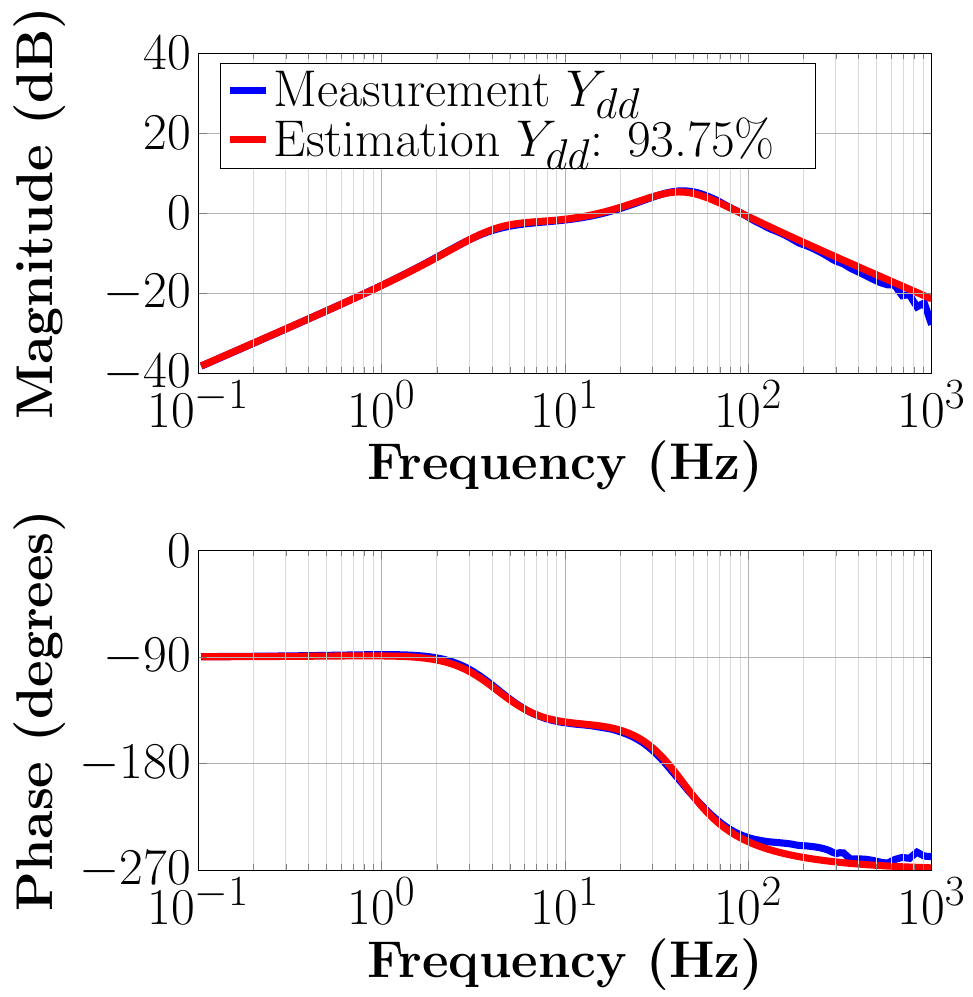}
        \label{fig:Ydd_SFRA}
    } 
    \subfloat[]{
        \includegraphics[width=0.2265\textwidth]{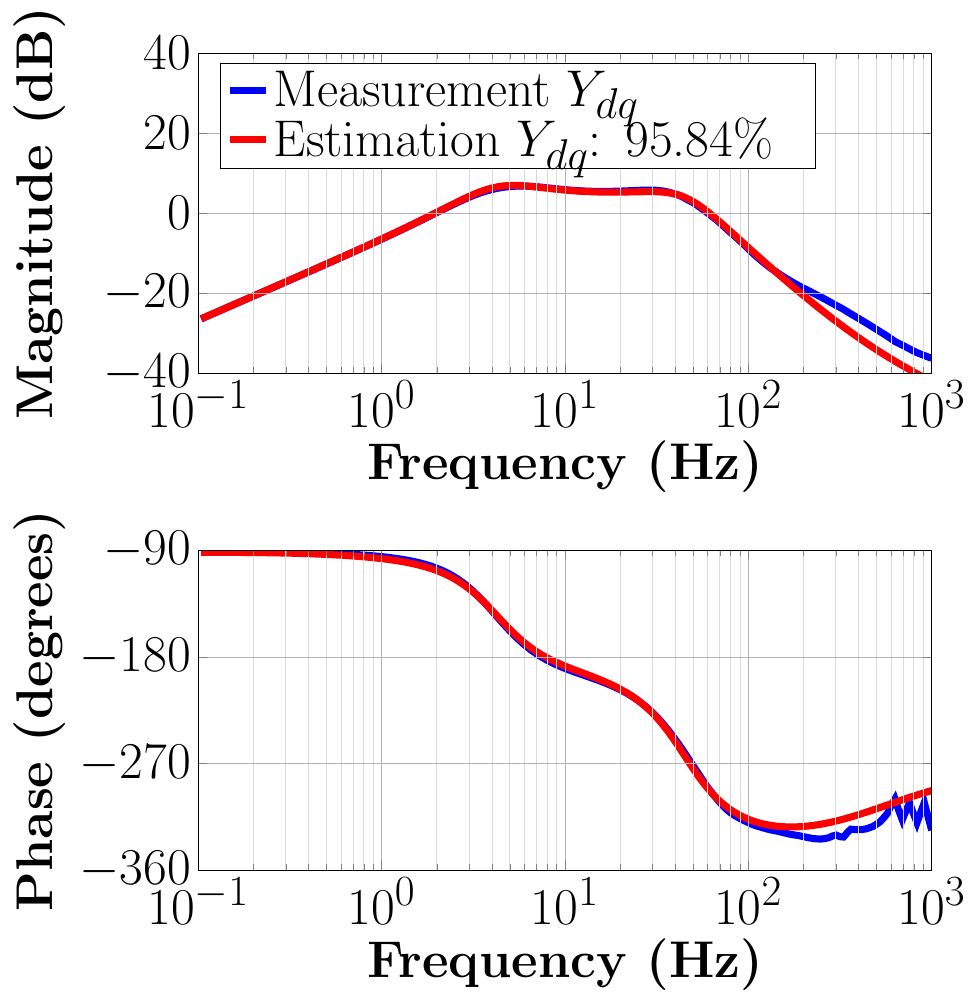}
        \label{fig:Ydq_SFRA}
    } 
    \subfloat[]{
        \includegraphics[width=0.2265\linewidth]{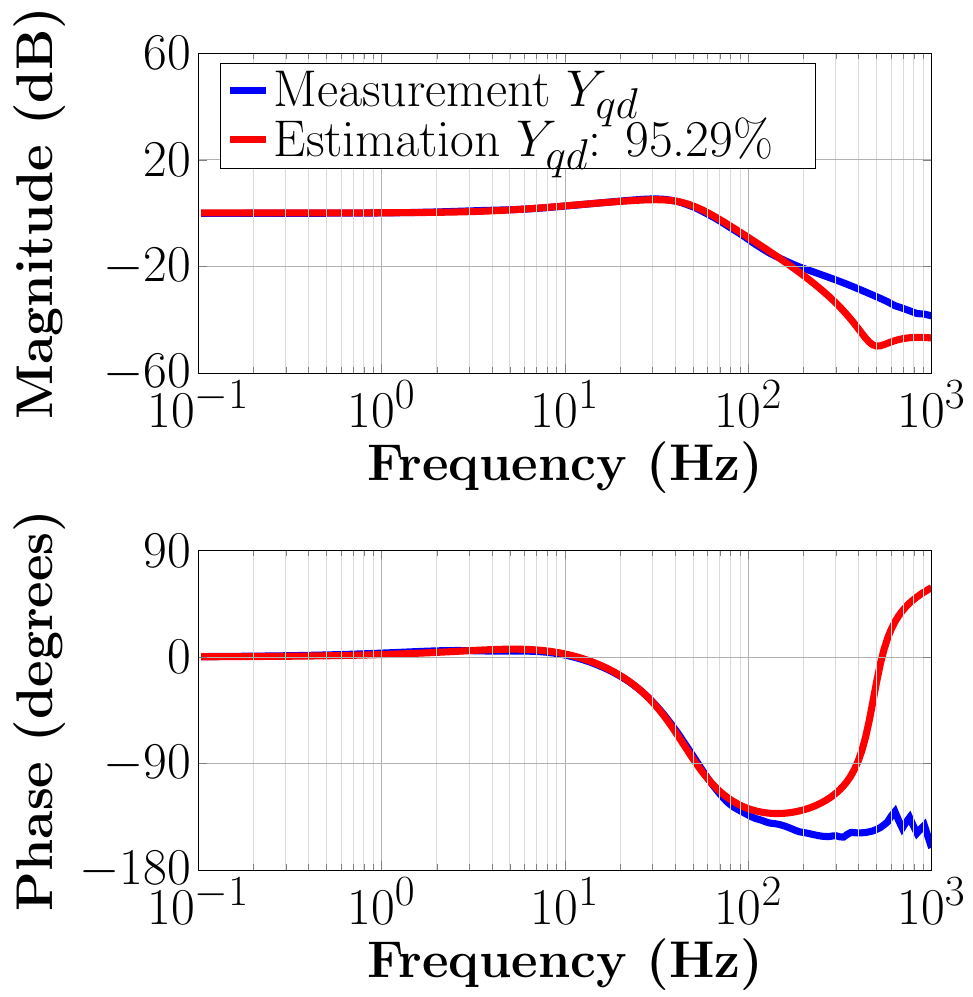}
        \label{fig:Yqd_SFRA}
    } 
   \subfloat[]{
        \includegraphics[width=0.2265\linewidth]{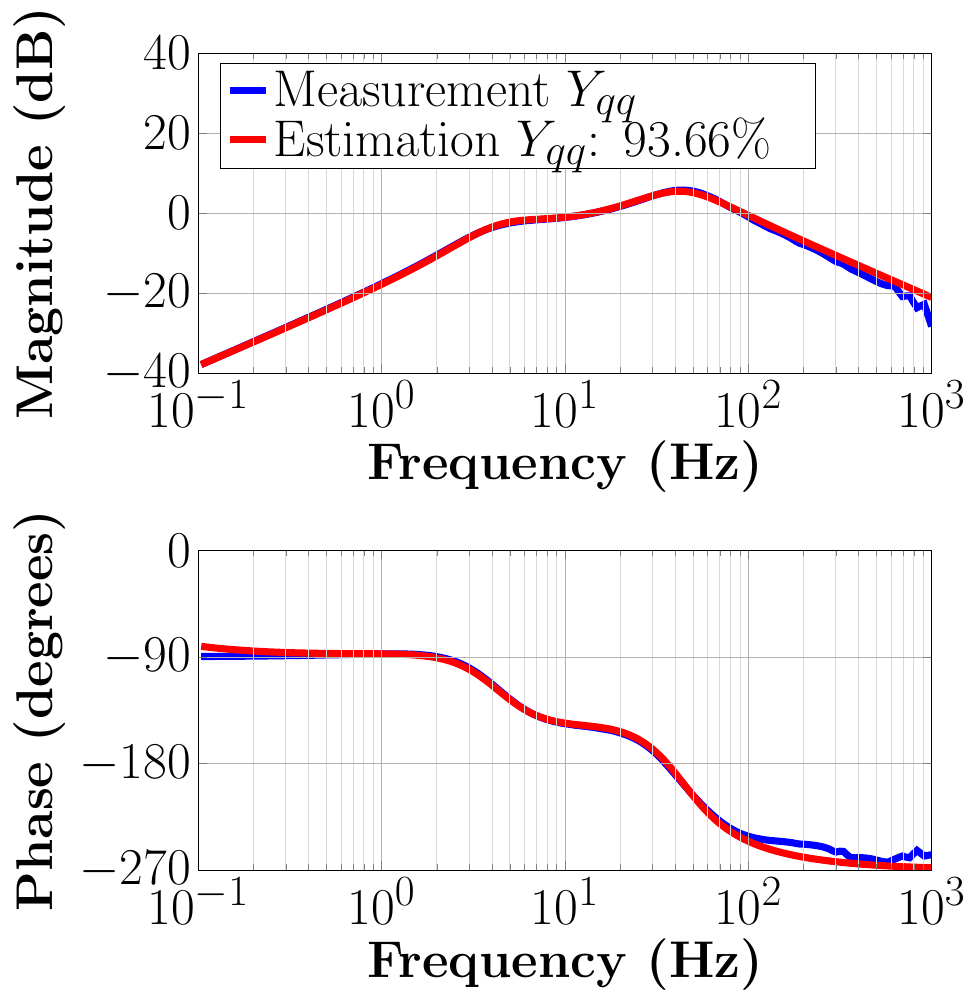}
        \label{fig:Yqq_SFRA}
    } \\

\vspace{-2mm}
\caption[CR]{Percentage of accuracy between the measurement frequency domain data and the estimation employing SFRA.} 
\vspace{-4mm}
\label{fig:SFRA_estimation}
\end{figure*}

In SFRA, a sinusoidal signal with a peak-to-peak amplitude of $0.1$V is injected separately into the grid’s $d$ and $q$ axis voltages. The frequency vector is considered to cover frequencies from $0.1$--$1000$ Hz, with the signal applied for two cycles at each frequency to ensure the system reaches a steady state after perturbation. This vector contains $100$ frequency points, meaning that $100$ injections are required to extract the admittance matrix according to \eqref{freq_scan}. To represent the transfer function in terms of a Bode plot, as shown in Fig. \ref{fig:admittance_plots}, \texttt{tfest} is applied to compute the transfer function from these discrete points, similar to the process used with SEM, but using frequency-domain data instead of time-domain. Fig. \ref{fig:SFRA_estimation} compares the simulated frequency-domain response with the estimated transfer function. The model is determined to be fourth-order, a value chosen iteratively by starting with a low-order model (one pole) and increasing the order until the percentage fit, measured by the NRMSE, exceeds 90\%. When this threshold is met, the simulated and estimated responses align closely, as shown in Fig.~\ref{fig:SFRA_estimation}. The mismatch between real measurements and the estimation in the $100$–$1000$Hz range can be addressed by increasing the system order, but this adds complexity with additional poles. The current order balances capturing essential dynamics with minimal complexity.

Fig. \ref{fig:admittance_plots} presents the Bode plots of the four elements of the admittance matrix, \(Y_{dd}\), \(Y_{dq}\), \(Y_{qd}\), and \(Y_{qq}\), identified using ERA, SEM, and SFRA. For the four transfer functions, ERA and SEM show an overlapping frequency response, indicating similar results due to their shared use of step responses (see Figs.~\ref{fig:vgd_step_change} and \ref{fig:vgq_step_change}). Both methods operate similarly, as they estimate transfer functions using time-domain data. However, SFRA shows a different response due to its sinusoidal excitation method. At frequencies above $100$ Hz, there is a mismatch between ERA and SEM with SFRA. As mentioned in Section \ref{s:comparison}, when examining frequency responses at high frequencies, SFRA provides more accurate results since is possible to excite the system at discrete frequency points, covering a large range of frequencies. Section~\ref{s:comparison} also notes that no constraints are imposed on the frequency vector, enabling focused analysis of either low or high frequencies as long as the system remains in equilibrium.  Consequently, above $100$ Hz, SFRA remains more accurate due to its high SNR, where the signal strength is significantly higher than sensor noise. Time-domain methods as ERA and SEM are more effective at low frequencies, but their SNR decreases as frequency increases, making it challenging to capture high-frequency dynamics accurately. 


\vspace{-2mm}
\vspace{-1mm}
\section{Conclusions}\label{s:conclusion}
\vspace{-1mm}
This paper presents $dq$ admittance identification methods using frequency-domain techniques like SFRA and time-domain methods such as ERA and SEM. We show that ERA and SEM can effectively match SFRA results. A comparison  highlights the differences in signal injection, limitations, advantages, and suitable scenarios. SFRA excels at higher frequencies (above $100$ Hz) by targeting specific points with minimal noise interference, but it is time-intensive due to the wide frequency spectrum analysis required. In contrast, ERA and SEM offer faster identification with just two step changes at the $dq$ voltage source, making them ideal for low-frequency studies ($1-100$ Hz) with lower computational demands.


\vspace{-4mm}
\section*{{Acknowledgment}}
\vspace{-2mm}
{This work has been supported by KAUST under Award No. ORFS-2022-CRG11-5021.}
\vspace{-5mm}
\bibliographystyle{ieeetr}
\bibliography{references}
\vfill

\end{document}